\documentclass[
 aps, superscriptaddress,
 amsmath,amssymb,
 reprint,  
]{revtex4-1}
\usepackage[utf8]{inputenc}
\usepackage{graphicx}
\usepackage{float}
\usepackage{enumitem}
\usepackage{bm}
\usepackage{bbold}
\usepackage{cases}
\usepackage{upgreek}
\usepackage[usenames, dvipsnames]{color}
\usepackage[colorlinks=true, citecolor=Maroon, linkcolor=Bittersweet, urlcolor=RawSienna]{hyperref}
\usepackage{hyperref}
\usepackage{braket}
\usepackage[normalem]{ulem}

\usepackage{textgreek}
\usepackage{gensymb}
\usepackage{dcolumn}

\newcommand \mc[1] { \mathcal{#1} }
\newcommand \dd[1]  { \!\!\textrm d{#1} \,}   
\newcommand \rmm[1]  { \textrm{#1} }

\newcommand \e[1] { \rmm{e}^{#1} }

\makeatletter
\def\@email#1#2{%
 \endgroup
 \patchcmd{\titleblock@produce}
  {\frontmatter@RRAPformat}
  {\frontmatter@RRAPformat{\produce@RRAP{*#1\href{mailto:#2}{#2}}}\frontmatter@RRAPformat}
  {}{}
}%
\makeatother
\begin{document}

\preprint{AIP/123-QED}

\title{Nonequilibrium relaxation exponentially delays the onset of quantum diffusion}

\author{Srijan Bhattacharyya}
\affiliation{Department of Chemistry, University of Colorado Boulder, Boulder, CO 80309, USA}

\author{Thomas Sayer}
\affiliation{Department of Chemistry, University of Colorado Boulder, Boulder, CO 80309, USA}
\affiliation{Department of Chemistry, Durham University, Durham DH1 3LE, United Kingdom}

\author{Andr\'{e}s Montoya-Castillo}
\homepage{Andres.MontoyaCastillo@colorado.edu}
\affiliation{Department of Chemistry, University of Colorado Boulder, Boulder, CO 80309, USA} 


\date{\today}

\begin{abstract}
Predicting the exact many-body quantum dynamics of polarons in materials with strong carrier-phonon interactions presents a fundamental challenge, often necessitating one to adopt approximations that sacrifice the ability to predict the transition from nonequilibrium relaxation to thermodynamic equilibrium. Here, we exploit a recent breakthrough that generalizes the concept of memory beyond its conventional temporal meaning to also encompass space. Specifically, we leverage our discovery that the dynamics of observables in systems with local couplings satisfy Green's functions with kernels that are local in time and space. This enables us to employ the dynamics of small lattices over short times to predict the dynamics of thermodynamically large lattices over arbitrarily long timescales while circumventing the deleterious impacts of finite-size effects. We thus interrogate the \textit{exact} nonequilibrium formation and migration of polarons in one- (1D) and two-dimensional (2D) systems, revealing that their motion approaches diffusive transport only asymptotically in time and system size. We also compare transport in 1D and 2D systems to investigate the effect of dimension in polaron migration physics, illustrating how energy variations can cause localization---a phenomenon observable via current microscopy experiments.
\end{abstract}

\maketitle

\textbf{\textit{Introduction}} -- Predicting and elucidating transport in materials is crucial for their practical use in various technologies~\cite{balhorn_closing_2022, bhat_computational_2023}, such as polymers in electronics~\cite{fratini_charge_2020, tjhe_non-equilibrium_2024} or transition metal oxides in photocatalysis~\cite{pastor_electronic_2022, wang_giant_2023}. In these materials, polarons---an electronic excitation (e.g., a charge or exciton) and the deformation it causes in the surrounding material---serve as the primary energy carriers~\cite{emin_polarons_2012, franchini_polarons_2021}. Understanding how a material's microscopic properties determine polaron transport requires the ability to measure and simulate their quantum dynamics with controllable accuracy over experimentally relevant system sizes and length scales. Recent advances in microscopy enable the spatiotemporal measurement of polaron transport~\cite{zhu_ultrafast_2019, ginsberg_spatially_2020}, revealing how energy flows in materials spanning organic polymers~\cite{delor_imaging_2020}, quantum dots~\cite{yoon_direct_2016}, perovskites~\cite{guo_long-range_2017}, and transition metal dichalcogenides~\cite{chernikov_exciton_2014}. But simulation methods have not kept pace: current theories fail to predict energy flow even in widely adopted models (e.g., Holstein~\cite{holstein_studies_1959}, Su-Schrieffer–Heeger~\cite{su_solitons_1979}, Fr\"ohlich~\cite{frohlich_electrons_1954}) over experimentally relevant system sizes and timescales. 

Essentially, quantum dynamical methods are often beset by astronomically expensive computational scaling. To lower these costs, one may invoke projection operator techniques~\cite{grabert_projection_1982}, which yield a low-dimensional equation of motion---a generalized master equation (GME)~\cite{nakajima_quantum_1958, zwanzig_ensemble_1960, mori_transport_1965}---for a few observables. GMEs are broadly applicable to problems ranging from structural polymer relaxation~\cite{schweizer_microscopic_1989, chatterjee_calculation_1994}, to vitrification~\cite{dyre_master-equation_1987, janssen_microscopic_2015, nandi_role_2017, garrahan_aspects_2018, liu_dynamics_2021, bidhoodi_long-time_2023}, protein folding~\cite{ayaz_non-markovian_2021, cao_advantages_2020, vroylandt_likelihood-based_2022, cao_integrative_2023, dominic_building_2023, dominic_memory_2023, dalton_fast_2023}, and quantum decoherence~\cite{maniscalco_non-markovian_2006, barnes_nonperturbative_2012, de_vega_dynamics_2017, campaioli_quantum_2024}, and have reduced the cost of predicting charge transfer~\cite{pfalzgraff_nonadiabatic_2015} and the biomolecular dynamics of protein folding~\cite{dominic_building_2023} by multiple orders of magnitude. These savings are possible because the non-Markovian description of the reduced dynamics (memory kernel) has a lifetime often shorter than the relaxation time of the dynamical variables one seeks to simulate. GMEs thus enable one to simulate reduced dynamics to arbitrarily long times as long as the reference calculation used to construct the memory kernel is at least as long as its lifetime. Yet, while GMEs have been developed to predict polaron transport in the Holstein model~\cite{yan_theoretical_2019}---in principle enabling researchers to access the long-time limit of polaron dynamics on a finite lattice---the lattice boundaries cause particles to reflect off them (or interfere) in finite-chain (periodic) systems. These finite-size effects poison simulations past this reflection time and have thus prevented the systematic interrogation of how nonequilibrium excitations relax, approach equilibrium, and transition from far-from-equilibrium non-diffusive transport to near-equilibrium diffusion. The only solution therefore appears to perform quantum dynamics simulations of polaron formation and transport in larger systems. However, the nonlinear scaling of direct and traditional GME-based simulations with system size prevents one from reaching results in the thermodynamic limit of infinite size.

We solve this problem by exploiting our recent GMEs for lattices with long-range order and local interactions that exhibit \textit{finite memory in time and space} which can be constructed from small-lattice, short-time simulations~\cite{bhattacharyya_space-local_2024}. This enables us to reach the thermodynamic limit free of finite-size effects. We leverage these novel GMEs to interrogate nonequilibrium polaron flow in the dispersive Holstein model~\cite{mahan_many-particle_2000, yan_theoretical_2019} in one-dimensional (1D) and 2D lattices. Our simulations \textit{reveal that diffusion onsets exponentially slowly (asymptotically) in the limit of large homogeneous systems}. Contrary to expectations of the contrasting behavior of transport in one versus multiple dimensions, we also show that capturing polaron transport in the dilute limit in 1D determines transport in higher dimensions. We also demonstrate how energy scale variations can localize polaron motion along specific directions. 

\textbf{\textit{System}} -- The Holstein Hamiltonian is the cornerstone of small polaron physics~\cite{holstein_studies_1959}. It encodes how carriers interact with their local lattice to form polarons, and has been applied to a wide variety of chemical systems, including organic crystals~\cite{cheng_unified_2008}, polymers~\cite{qarai_understanding_2021}, transition metal oxides~\cite{andreoni_small_2020}, and nanomaterials~\cite{mousavi_effects_2012} to unravel carrier motion. While the original Holstein model~\cite{holstein_studies_1959} considers local coupling to a single optical phonon mode, we focus on its dispersive counterpart, where a charge carrier couples to a continuum of phonon modes, better mimicking the broad spectrum of carrier-phonon couplings revealed by atomistic simulations~\cite{nematiaram_modeling_2020}. 

The dispersive Holstein Hamiltonian with $N$ lattice sites is defined as
\begin{equation} \label{eq:hamiltonian}
    \hat{H} = \sum_{i}^{N}\epsilon_i \hat{a}_i^\dag \hat{a}_i + \sum_{\langle ij \rangle }^{N} v_{ij} \hat{a}_i^\dag \hat{a}_j 
   +  \sum_{i}^{N}\big[ \hat{H}_{i}^{\mathrm{ph}} + \hat{H}_{i}^{\mathrm{el-ph}}\big],
\end{equation}
where $\hat{a}_i^\dag$ ($\hat{a}_i$) is a fermionic creation (annihilation) operator, the index $i$ labels lattice sites, $\epsilon_i$ is the carrier energy on the $i$th site, $v_{ij}$ is the hopping integral, and $\langle ij \rangle$ denotes that $v$ only connects nearest neighbors. Each site couples to local harmonic phonons described by $\hat{H}_{i}^{\mathrm{ph}} =  \sum_{\alpha }\omega_{i\alpha} \hat{b}_{i, \alpha}^\dag \hat{b}_{i, \alpha}$ with frequencies $\omega_{i, \alpha}$. $\hat{b}_{i, \alpha}^\dag$ ($\hat{b}_{i, \alpha}$) is the bosonic creation (annihilation) operator, and $\alpha$ indexes the phonon modes. The carrier-phonon coupling, $ \hat{H}_{i}^{\mathrm{el-ph}} = \sum_{\alpha} g_{i, \alpha}  \hat{a}_i^\dag \hat{a}_i (\hat{b}_i^\dag + \hat{b}_i)$, modulates the on-site energies, enabling polaron formation. The site-specific spectral densities, $ J_i(\omega)= \pi \sum_{\alpha} g_{i,\alpha}^2  \delta(\omega -\omega_{i\alpha})$, characterize local carrier-phonon couplings. Consistent with previous studies~\cite{yan_theoretical_2019, bhattacharyya_anomalous_2024}, we adopt the dilute limit (one charge manifold) in homogeneous systems where $\epsilon_i=0$, $v_{ij} = v$, $r_0 = $ 5~\AA~is the intersite lattice distance, and $J_i (\omega) = J(\omega)= \dfrac{ \eta \gamma \omega}{\omega^2 + \gamma^2} $ are of Ohmic-Debye form encoding condensed phase dissipation~\cite{weiss_quantum_2012}. Here, $\eta/2$ is the reorganization energy and $1/\gamma$ is the phonon decorrelation time. 

When tackling the dynamics of the dispersive Holstein model with numerically exact techniques, like the hierarchical equations of motion (HEOM)~\cite{shi_efficient_2009, song_time_2015}, the computational cost of solving the many-body dynamics of sufficiently large systems becomes prohibitive. This challenge intensifies when investigating nonequilibrium relaxation because of its slow convergence with system size~\cite{bhattacharyya_anomalous_2024}. Inspired by recent experiments that track excitation density in space~\cite{chernikov_exciton_2014, yoon_direct_2016, guo_long-range_2017, zhu_ultrafast_2019, delor_imaging_2020, ginsberg_spatially_2020}, we confront the cost issue by adopting a GME that exclusively tracks the time-dependent lattice-site populations after initial carrier injection (or photogeneration). Hence, we are interested in a correlation matrix, $C_{ij}(t)$, that tracks the time-dependent probability of finding a carrier on site $i$ after an initial carrier injection on site $j$ (see Methods section).

To analyze the site-resolved non-equilibrium population dynamics we compute the polaron's mean-squared displacement, $MSD = \sum_k d_k^2 C_{k}(t)$, which encodes its diffusion constant, 
\begin{equation}\label{eq:diffusion}
    D =  \lim_{t \to \infty} \frac{1}{2 N_d} \frac{\mathrm{d}MSD} {\mathrm{d}t}.
\end{equation}
Here, the translational invariance of the homogeneous lattice means that one can collapse the pair of indices into one, $C_{k= i-j \mod N}(t) = C_{ij}(t)$, and $N_d$ denotes the lattice dimension. For $N_d > 1$, the indices in $C_{ij}(t)$ become vector-valued and of dimension equal to $N_d$. 

\begin{figure}[t]
\begin{center} 
\vspace{-3pt}
    \resizebox{.4\textwidth}{!}{\includegraphics[trim={5pt 5pt 5pt 5pt},clip]{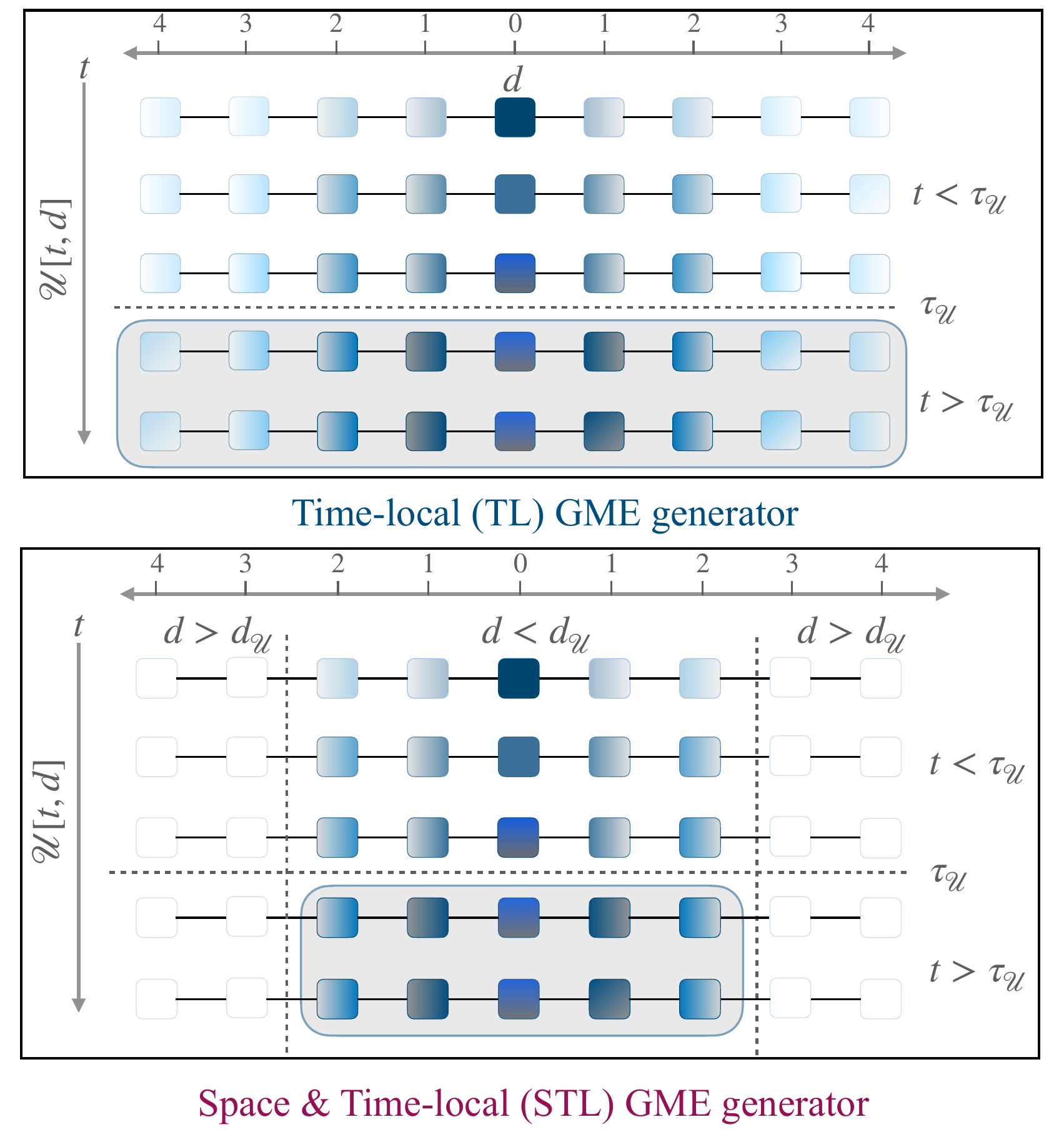}}
    
\vspace{-3pt}
\end{center}
\caption{\label{fig:schematic} Schematic of the non-Markovian generator $\mathcal{U}$ in time (vertical) and space (horizontal) axes. The color intensity signifies the magnitude of each element. \textbf{Top}: Conventional time-local GME where all elements become constant after $ \tau_{\mathcal{U}} $.  \textbf{Bottom}: spatial truncation in our space- and time-local GME after the characteristic distance, $d_{\mathcal{U}}$, beyond which all elements become insignificant and can be set to zero. }
\vspace{-5pt}
\end{figure}

\textbf{\textit{Argument for Space-Local Memory}} -- 
Here we provide a unified view of our recently developed space- and time-local (STL) GME~\cite{bhattacharyya_space-local_2024}, which offers a convenient framework to interrogate the dynamics of dispersive Holstein polarons. We further elicit connections between our STL-GME framework and the powerful language of spatiotemporal Green's functions. 

\begin{figure*}
\begin{center} 
\vspace{-3pt}
    \resizebox{.9\textwidth}{!}{\includegraphics[trim={0pt 0pt 3pt 0pt},clip]{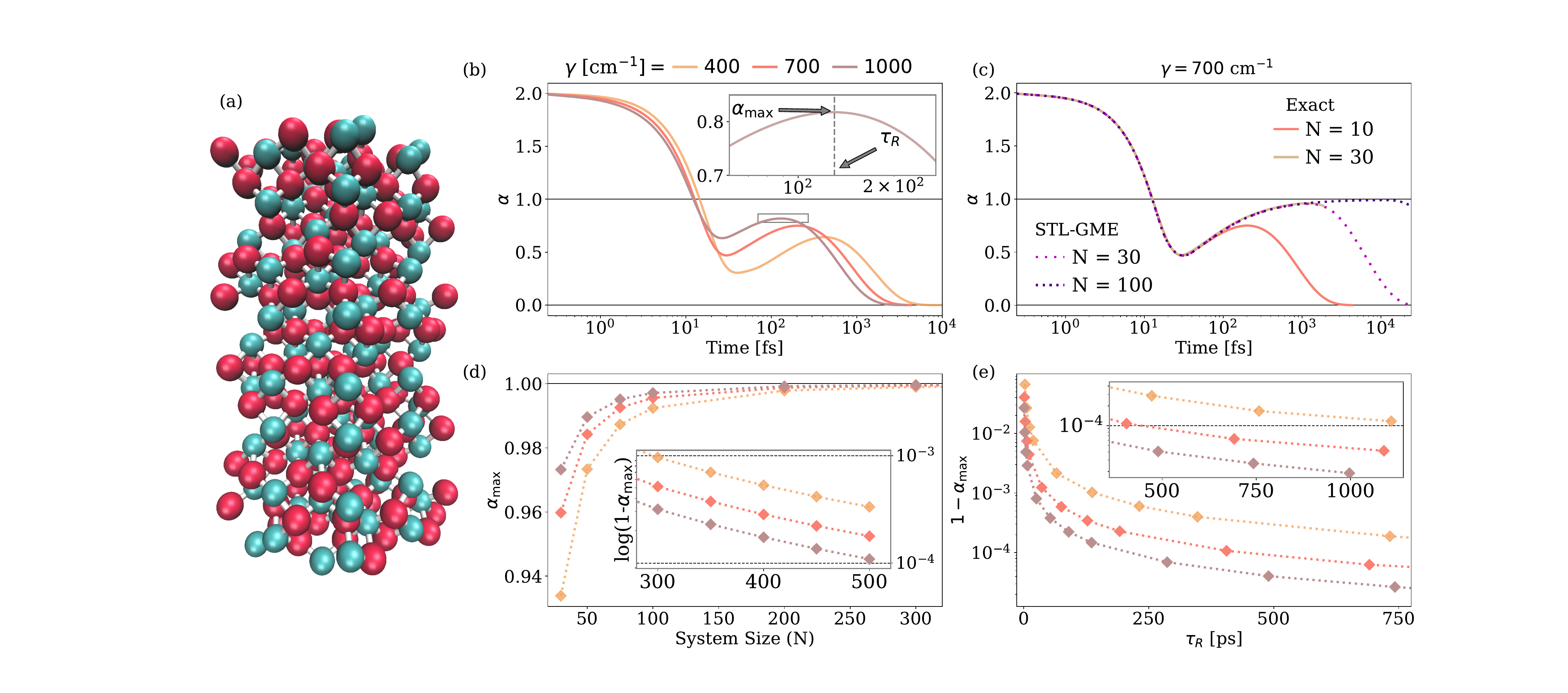}}
\vspace{-15pt}
\end{center}
\caption{\label{fig:iron_parameter} Dependence of the scaling exponent, $\alpha$, for polaron transport in the dispersive Holstein model with $v_{ij} = 322$ cm$^{-1}$, $\eta = 2984$ cm$^{-1}$, $T = 300 $ K, and $\gamma = [400, 700, 1000]$ cm$^{-1}$. \textbf{(a)} Schematic of the hematite crystal lattice (Fe in teal and O in red). \textbf{(b)} Variation of $\alpha$ as a function of $\gamma$. \textbf{Inset}: The maxima of $\alpha$ occurs at timescale $\tau_R$ after which $\alpha$ decreases due to finite-size. \textbf{(c)} STL-GME extension of a numerically exact 10-site polaron transport simulation to 30 and 100 sites. \textbf{(d)} Dependence of $\alpha_{\mathrm{max}}$ on system size. \textbf{Inset}: Demonstration of the linear scaling of $1 - \alpha_{\mathrm{max}}$ with system size in the large-system length limit.  \textbf{(e)} Log-scale plot of $1 - \alpha_{\mathrm{max}}$ with finite size-induced reflection time, $\tau_R$.  \textbf{Inset}: Log-scale plot of $1 - \alpha_{\mathrm{max}}$ as a function of large $\tau_R$.}
\vspace{-5pt}
\end{figure*}

We begin with the integrated time-local (TL) GME for $\bm{C}(t)$~\cite{sayer_compact_2023},
\begin{equation}\label{eq:tcl-GME}
    C_{i,j}(t+\delta t) = \sum_{k} \mathcal{U}_{i,k} (t) C_{k,j}(t), 
\end{equation}
where $ \bm{\mc{U}} (t)$ is the generator that describes its non-Markovian dynamics. In dissipative systems, $\bm{\mathcal{U}}(t)$ becomes constant after some time, denoted as the lifetime, $\tau_{\mathcal{U}}$. Hence, if one can compute reference dynamics up to $\tau_{\mathcal{U}}$, one can predict the long-time dynamics using Eq.~\ref{eq:tcl-GME} at a comparatively trivial computational cost. Figure ~\ref{fig:schematic}, top, illustrates that $\bm{\mathcal{U}}(t> \tau_{\mathcal{U}})$ stops changing, transforming Eq.~\ref{eq:tcl-GME} into a simple rate equation. Since $j$ and $i$ specify the position of the initial excitation and final measurement, one can rewrite the index dependence using the index difference, or relative position, $\mathbf{r}$ in systems with translational invariance: $C_{i,j}(t) \rightarrow C(\mathbf{r}, t)$. Similarly, the index dependence of $\bm{\mathcal{U}}(t)$ can be simplified to a relative position. Thus,
\begin{equation}\label{eq:tcl-GME-greens-function}
    {C}(\mathbf{r}, t+\delta t) = \int \dd{\bm{r}'}{\mc{U}}(\mathbf{r} - \mathbf{r}', t) {C}(\mathbf{r}', t).
\end{equation}
Here, $\mc{U}(\mathbf{s}, t)$ quantifies the strength of transitions separated by $\mathbf{s}$ at time $t$. Our insight is that for short-range couplings, $\mc{U}(\mathbf{s}, t)$ is local in space, i.e., $\mc{U}(\mathbf{s}, t)$ decays with increasing distance from the initial excitation $d = |\mathbf{s}|$.

This insight arises from a simple physical picture. Consider a quasiparticle moving with characteristic speed $\mathcal{V}$ in real or momentum space. By $t = \tau_{\mathcal{U}}$, the particle will have traveled $d_{\mathcal{U}} \sim \tau_{\mathcal{U}} \times \mathcal{V}$, such that $\mc{U}(|\mathbf{s}|> d_{\mathcal{U}}, t) \sim 0$ and does not affect the dynamics. Hence, as Fig.~\ref{fig:schematic}, bottom, illustrates, if $ \bm{\mathcal{U}}$ has finite memory \textit{in time}, it must also have finite memory \textit{in space}. Thus, constructing $\mathcal{U}(|\mathbf{s}| \leq d_{\mc{U}}, t \leq \tau_{\mc{U}})$ from the reference dynamics of an $N$-site lattice where $d_{\mathcal{U}} < N/2$ keeps it free of finite-size effects. We exploit this space-time locality to extend the dynamics of small lattices to thermodynamically large systems over arbitrarily long times. 

When coupled with numerically exact dynamics, such as those we employ here (HEOM), our STL-GME offers an accurate and efficient way to generate the dynamics of lattice-based Hamiltonians with local couplings, like the dispersive Holstein model~\cite{bhattacharyya_space-local_2024}. Specifically, our STL-GME can reduce the computational scaling of quantum dynamics simulations of lattice problems from polynomial (even exponential) in time and system size to quadratic scaling with a small prefactor that heavily suppresses the cost. In practice, this cost is negligible compared to the cost of the requisite small-system, short-time reference calculation. Thus, we employ short-time, small-lattice HEOM dynamics to construct our STL-GME for large systems over arbitrary times, thereby enabling us to bridge local nonequilibrium relaxation with global thermodynamic equilibration. 

\begin{figure*}
\begin{center} 
\vspace{-3pt}
    \resizebox{.9\textwidth}{!}{\includegraphics[trim={00pt 0pt 0pt 0pt},clip]{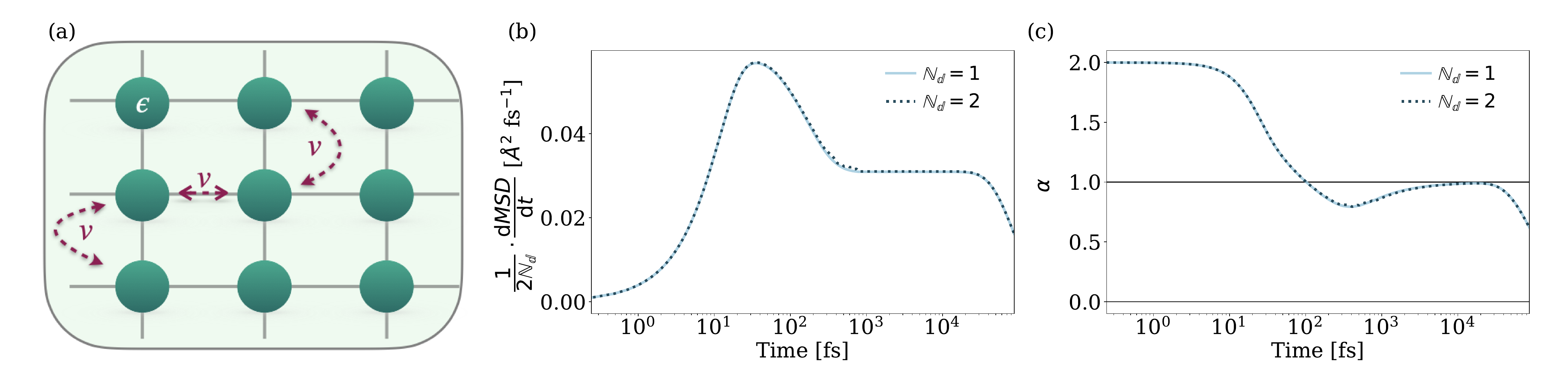}}
    \vspace{-15pt}
\end{center}
\caption{\label{fig:2dvs1d_polaron} Transport properties of dispersive Holstein polarons in 2D with $\eta = 500$ cm$^{-1}$, $\gamma = 41$ cm$^{-1}$, and $T = 300 $ K.\textbf{(a)} Homogeneous 2D lattice. \textbf{(b)} Comparison of $ \frac{1}{2 N_d}  \frac{\mathrm{d}MSD}{\mathrm{d}t}$. \textbf{(c)} Comparison for the scaling exponent $\alpha$ for a 2D $30 \times 30$-site lattice (solid line) and a 1D $30$-site lattice (dotted line) with $\epsilon_1 = 0$ cm$^{-1}$ and $v = 25$ cm$^{-1}$.}
\vspace{-5pt}
\end{figure*}

\textbf{\textit{Polaron transport}} -- The $MSD$ and its power law, $MSD \propto t^{\alpha}$, offer insights into transport mechanisms, with $\alpha < 1$ indicating subdiffusion, $\alpha = 1$ diffusion, and $\alpha > 1$ superdiffusion. While we have recently shown that homogeneous lattices exhibit subdiffusion for unexpectedly long times ($\sim20$ ps), our previous work was limited to a $\sim40$-site lattice, preventing us from reaching diffusion free from finite-size effects~\cite{bhattacharyya_anomalous_2024}. Our STL-GME's ability to tackle large systems over arbitrarily long times now enables us to investigate if, when, and how this anomalous polaron flow vanishes at the thermodynamic limit.

To address this question, we focus on hematite (Fig.~\ref{fig:iron_parameter} (a))---a well-known small polaron-forming system with applications in photocatalytic water splitting and as a photoanode material for solar-to-fuel conversion~\cite{carneiro_excitation-wavelength-dependent_2017}---with Hamiltonian parameters $v_{ij} = 322$ cm$^{-1}$, $\eta = 2984$ cm$^{-1}$~\cite{ahart_polaronic_2020}. We choose to study a set of values $\gamma = [400, 700, 1000]$ cm$^{-1}$, consistent with atomistic simulations~\cite{li_excitation-wavelength-dependent_2023}. Figure~\ref{fig:iron_parameter} (b) shows that, for a 10-site system, $\alpha$ initially displays ballistic behavior ($\alpha = 2$), then decreases to a subdiffusive regime that rises toward but does not reach diffusive transport ($\alpha = 1$), and then plummets to zero as finite-size effects emerge. The width and position of the subdiffusive peak vary with the characteristic dissipative speed of the phonon environment, $\gamma$: polarons coupled to higher-frequency phonons (large $\gamma$) exhibit a more contracted relaxation before the onset of finite-size effects, consistent with a fast-relaxing lattice that promotes polaron motion~\cite{bhattacharyya_anomalous_2024}. Yet, the dispersive Holstein polaron is expected to display diffusive behavior, consistent with calculations of its diffusion constant via the equilibrium Green-Kubo formulation~\cite{takahashi_tensor-train_2024, bhattacharyya_mori_2024}.  This begs the question: how large of a dispersive Holstein system is required to observe diffusive polaron transport after a nonequilibrium excitation? Figure~\ref{fig:iron_parameter} (c) shows our results for $\gamma = 700$ cm$^{-1}$. For 10 sites, $\alpha_{\mathrm{max}} \sim 0.78$ occurs around $\sim 200$ fs, while for 30 sites  $\alpha_{\mathrm{max}} \sim 0.95$ at $\sim 2$ ps. With our STL-GME, we find that for 100 sites,  $\alpha_{\mathrm{max}} \sim 0.99$ at  $\sim 20$~ps. 

Does $\alpha_{\rm max}$ reach 1 before the infinite lattice limit? We interrogate lattices of up to 1500~sites over timescales of 1.5~ns in Fig.~\ref{fig:iron_parameter} (d). Surprisingly, $\alpha_{\rm max}$ approaches unity for all $\gamma$ values \textit{exponentially slowly} as a function of system size. To wit, the logarithmic behavior of $1 - \alpha_{\mathrm{max}}$ in the limit of large lattices scales linearly across all values of $\gamma$. Similarly, Fig.~\ref{fig:iron_parameter} (e) shows that $1 - \alpha_{\mathrm{max}}$ starts to display exponential behavior (linear in log scale) at long $\tau_R$ (which indicates the time when the system is closest to diffusive motion). In the inset of Fig.~\ref{fig:iron_parameter} (e), we confirm this asymptotic behavior by plotting the same quantity in the long time limit ($500-1000$~ps). Importantly, the speed of the phonon bath (i.e., the magnitude of $\gamma$) dictates the system size, $N$, for which one can see the asymptotic behavior of $\alpha_{\mathrm{max}}$ as a function of $\tau_R$. For example, when $\gamma = 400$~cm$^{-1}$, the asymptotic behavior is already evident for system sizes of $N$~= 600-1000 sites, whereas when $\gamma = 1000$~cm$^{-1}$, the asymptotic behavior only starts to arise for system sizes of $N$~= 1000-1500 sites. Hence, nonequilibrium transport simulations \textit{only} reveal diffusion ($\alpha = 1$) \textit{asymptotically in time and in formally infinite systems.}

\begin{figure*}
\begin{center} 
\vspace{-3pt}
    \resizebox{.9\textwidth}{!}{\includegraphics[trim={00pt 0pt 0pt 0pt},clip]{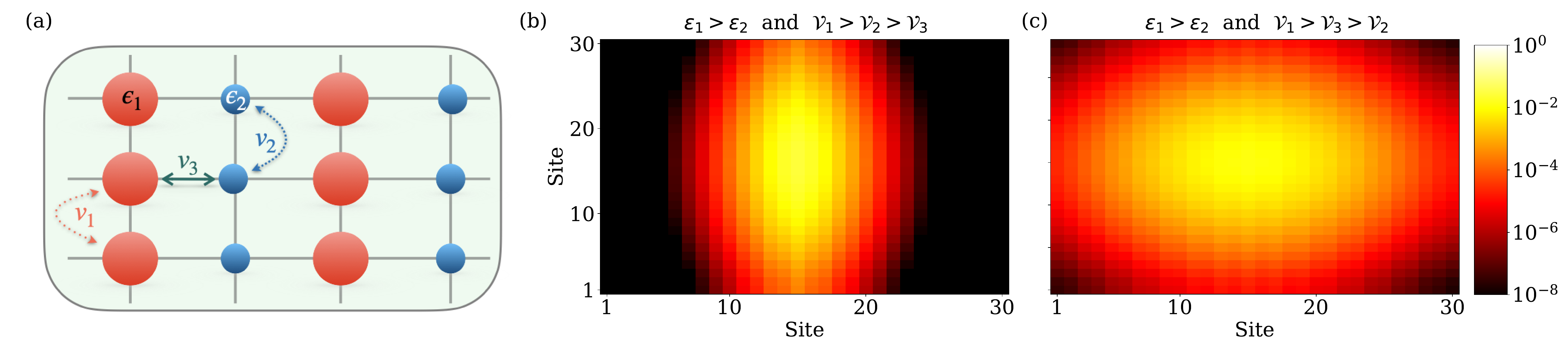}}
    \vspace{-15pt}
\end{center}
\caption{\label{fig:periodic_lattice} Polaron transport in 2D $30 \times 30$-site dispersive Holstein lattices at $t \sim 40$ with $\eta = 500$ cm$^{-1}$, $\gamma = 41$ cm$^{-1}$, and $T = 300 $ K. \textbf{(a)} Schematic of the periodic lattices with two distinct lattice points: red and blue sites. \textbf{(b)} Polaron density (when initiated on a red site) where $\epsilon_1 = 30$ cm$^{-1}$, $\epsilon_2 = 10$ cm$^{-1}$, $v_1 = 25$~cm$^{-1}$, $v_2 = 0.8 v_1$ and $v_3 = 0.3 v_1$.  \textbf{(c)} Polaron density (when initiated on a red site) where $\epsilon_1 = 30$ cm$^{-1}$, $\epsilon_2 = 10$ cm$^{-1}$, $v_1 = 25$ cm$^{-1}$, $v_2 = 0.3 v_1$ and $v_3 = 0.8 v_1$.}
\vspace{-17pt}
\end{figure*}

\textbf{\textit{Transport in 2D}} --
Having interrogated nonequilibrium polaron flow in 1D, we extend our analysis to higher dimensions, where transport characteristics can undergo dramatic transformations~\cite{isichenko_percolation_1992, piraud_quantum_2013, manzano_quantum_2016}. We generalize the STL-GME to $N$D lattices and specifically focus on 2D lattices, which serve as a valuable platform for direct comparison with cutting-edge microscopy experiments~\cite{chernikov_exciton_2014, yoon_direct_2016, guo_long-range_2017, zhu_ultrafast_2019, delor_imaging_2020, ginsberg_spatially_2020, tulyagankhodjaev_room-temperature_2023}. Our application of STL-GME allows us to capture the exact quantum dynamics of polaron formation and transport in unprecedentedly large 2D lattices (up to a $30 \times 30$-site configuration). We set parameters $v = 25$ cm$^{-1}$, $\eta = 500$ cm$^{-1}$, and $\gamma = 41$ cm$^{-1}$, consistent with small polaron-forming organic crystals~\cite{troisi_prediction_2007}. For details on the 2D STL-GME, see the Methods section. 

Whereas in 1D there is only one path for a polaron to reach a particular site, more paths become available in higher dimensions. For many systems, this only affects the transport coefficient by a dimensional factor, but we might expect larger, qualitative differences in the presence of many-body interactions. We work in the dilute limit with no direct carrier-carrier interactions, but the carrier can (self) interact indirectly via the phonon modes. It is therefore unclear whether 1D transport captures all higher dimensions for the dispersive Holstein model. We interrogate which limit dilute polaron transport satisfies on a homogeneous lattice (Fig.~\ref{fig:2dvs1d_polaron}(a)). Our results reveal that characteristic transport metrics, such as the dimensionality-renormalized $\frac{\mathrm{d}MSD}{\mathrm{d}t}$ (which reports on the diffusion constant) and the scaling exponent, $\alpha$, quantitatively agree across 1D and 2D lattices, as shown in Fig.~\ref{fig:2dvs1d_polaron}(b) and (c), respectively. Hence, we find that one can employ a dramatic simplification: an $N$-dimensional outer product of the 1D lattice dynamics can generate nonequilibrium polaron dynamics in $N$D. Yet, we note that the resulting polaron density in 2D does indeed differ when obtained by direct simulation rather than through the 1D outer product. However, this difference does not quantitatively affect the transport results because, first, the discrepancy is orders of magnitude smaller than the populations, second, it diminishes over time (see Supplementary information) and, third, the deviation takes both positive and negative values and therefore largely cancels when averaging around the solid angle. We, therefore, obtain the nuanced result that while dimensionality-renormalized $\frac{\mathrm{d}MSD}{\mathrm{d}t}$ and $\alpha$ reveal no significant differences for polaron transport in 1D and 2D, even in the early non-Markovian regime, the outer product of the 1D result does not exactly reproduce the true 2D simulation.

Having successfully simulated polaron motion in a homogeneous 2D lattice, we turn to periodic lattices like that shown in Fig.~\ref{fig:periodic_lattice}(a). These 2D lattices introduce greater complexity and allow us to assess the effects of varying energy scales on polaron transport---an aspect that is particularly relevant for chemical systems and materials design~\cite{noriega_chain_2013, akselrod_visualization_2014, guzelturk_visualization_2021, buizza_polarons_2021, liu_atomic-scale_2023, tulyagankhodjaev_room-temperature_2023}. We consider a lattice with two types of sites with different on-site energies $(\epsilon_1 \neq \epsilon_2)$ and different couplings connecting equivalent $(v_1$ and $v_2)$ versus distinct ($v_3$) sites. Figure~\ref{fig:periodic_lattice}(b) shows the polaron spread when $\epsilon_1 > \epsilon_2$ and $v_1 > v_2 > v_3$ at $t \approx 40$~ps. The maximal polaron spread along the vertical axis shows that a comparatively small $v_3$ localizes the polaron along the horizontal axis. Instead, when $v_1 > v_3 > v_2$, there is no localization along (half the columns of) the vertical axis because charge can travel along the faster ($v_1$) direction and move horizontally ($v_3$), bypassing the lower coupling ($v_2$) pathway. Fig.~\ref{fig:periodic_lattice}(c) shows that these parameters produce an elliptical distribution ($v_1>v_3$) at long times, with no evidence of the variegated underlying energy landscape. By accessing polaron flow in a 2D lattice over unprecedentedly long timescales (120 ps), we thus quantify the dynamics over the full equilibration timescale (see Supplementary information).

\textbf{\textit{Outlook}} -- We interrogated the transition of nonequilibrium polaron formation and relaxation from the anomalous transport regime into the diffusive regime. Our study revealed that diffusion onsets asymptotically slowly as a function of lattice size and time in homogeneous lattices. We further elucidated the influence of dimensionality on polaron transport, demonstrating that nonequilibrium relaxation in uniform 1D lattices can quantitatively capture quantum transport behaviors observed in higher-dimensional systems. Moreover, we found that discrepancies in polaron density between 2D lattices and those approximated from 1D configurations diminish over time. Our simulations of polaron dynamics in periodic lattices revealed interesting localization phenomena, illustrating how energy landscapes can direct polaron motion asymmetrically across a 2D surface. 

Our study of polaron relaxation in 1D and 2D represents a crucial advance, providing a tool that can bridge theoretical simulation with experimental observations of polaron formation, localization, and directed motion. This enables a new route to investigate the mechanism of polaron formation and transport in, for example, 2D nanomaterial heterostructures~\cite{kang_holstein_2018} and anisotropic inorganic crystals such as mixed metal oxides~\cite{carneiro_excitation-wavelength-dependent_2017}. 

Beyond nonequilibrium polaron physics, we found that our recently developed STL-GME~\cite{bhattacharyya_space-local_2024} functions as a Green's function that enables one to harness the dynamics of small lattices with local couplings over short timescales to predict the relaxation of thermodynamically large systems across arbitrary timescales. This method is broadly applicable and offers new and efficient avenues for investigating spin relaxation in organic spintronics devices~\cite{schott_polaron_2019}, qubit crosstalk in quantum computing~\cite{zhao_quantum_2022}, topological magnons in ferromagnetic systems~\cite{mcclarty_topological_2022}, and transport phenomena such as thermal conduction~\cite{xia_unified_2023}. We have therefore laid the groundwork for future explorations into the complex many-body dynamics of materials over length and timescales that are directly comparable to experiments, in addition to taking a significant step forward in the theoretical understanding of polaron physics.

\vspace{-12pt}
\section*{Acknowledgments}
\vspace{-5pt}
A.M.C.~acknowledges the start-up funds from the University of Colorado Boulder for partial support of this research. Acknowledgment is made to the donors of the American Chemical Society Petroleum Research Fund for partial support of this research (No.~PRF 66836-DNI6). A.M.C.~acknowledges the support from a David and Lucile Packard Fellowship for Science and Engineering. S.B.~acknowledges the Marion L. Sharrah Fellowship for partial support of the study. T.S.~is the recipient of an Early Career Fellowship from the Leverhulme Trust. We thank Prof.~Qiang Shi for sharing his HEOM code with us. This work utilized the Alpine high-performance computing resource at the University of Colorado Boulder. Alpine is jointly funded by the University of Colorado Boulder, the University of Colorado Anschutz, Colorado State University, and the National Science Foundation (award 2201538).

\section*{Data Accessibility}
\vspace{-4pt}
The data supporting our study's findings are available from the corresponding author upon reasonable request.
\bibliography{references.bib}

\clearpage

\setcounter{section}{0}
\setcounter{equation}{0}
\setcounter{figure}{0}

\renewcommand{\theequation}{M\arabic{equation}}
\renewcommand{\thefigure}{M\arabic{figure}}

\section*{Methods}
\vspace{-8pt}
\subsection{Details of Exact dynamics} \label{HEOM-details}
\vspace{-5pt}
We have used the hierarchical equations of motion (HEOM)~\cite{shi_efficient_2009} as an exact solver to perform the dynamics of the dispersive Holstein system at finite temperatures. HEOM maps the dissipative environment in an open quantum system to auxiliary density matrices (ADMs) and solves a set of coupled differential equations for all ADMs at each time step. After integrating the environmental bosonic (phonon) degrees of freedom, HEOM predicts the electronic system's reduced density matrix, $\text{Tr} \left[ \hat{a}^\dagger_i \hat{a}_i\e{-\rmm{i}\mathcal{L}t}  \hat{\rho_{j}}(0)\right]$, where $j$ index denotes the initial lattice site where nonequilibrium relaxation begins, $i$ denotes the site on which one measures the polaron density at time $t$ with, and $\hat{\rho_j}(0) = \hat{a}_j^{\dagger}\hat{a}_j  \hat{\rho}^{\rm{ph}}$. Here, $\hat{\rho}^{\rm{ph}} = e^{-\beta \hat{H}^{\rm ph}}/\mathrm{Tr}[e^{-\beta \hat{H}^{\rm ph}}]$. We construct the population correlation matrix $\bm{C} (t)$ with this HEOM output for our further calculation with GME. 

We converged all HEOM calculations with respect to the hierarchical depth $L$, number of Matsubara frequencies $K$, and timestep $dt =0.25$ fs. We employed dynamic filtering in our simulations~\cite{shi_efficient_2009}, setting $\delta = 10^{-7}$ atomic units at each timestep of our HEOM simulations, and the n-particle approximation to reduce the computational cost of the HEOM simulations~\cite{song_time_2015}. To ensure convergence, we calculate the quantity $\frac{1}{N^2 N_{\rm tsteps}}{||C_{\rm HEOM} (p_1) - C_{\rm HEOM} (p_2)||_2}$ in the dynamics arising from two difference sets of parameters $\{p_1, p_2\}$. The difference is normalized by the total number of points $N_{\rm tsteps}$ and the dimension of the correlation matrix $N^2$. This allowed us to converge the HEOM reference dynamics with respect to $L$, $K$, and $dt$. For Fig.~\ref{fig:iron_parameter}, our converged parameters are $L=10$ (threshold $2 \times 10^{-7}$) and $K=1$ (threshold $1 \times 10^{-5}$), while for Figs.~\ref{fig:2dvs1d_polaron} and \ref{fig:periodic_lattice}, $L = 26$ (threshold $1 \times 10^{-9}$). Because HEOM becomes prohibitively expensive for 2D simulations, we employ the high-temperature approximation, $K = 0$.

\vspace{-8pt}
\subsection{Construction of GME framework} \label{GME-framework}
\vspace{-5pt}

To construct our STL-GME~\cite{bhattacharyya_space-local_2024}, we first need to build the TL-GME generator $\bm{\mathcal{U}}(t)$ from our small-lattice short-time HEOM dynamics, $C_{\rm ref}(t)$:
\begin{equation}\label{def-u}
    \bm{\mathcal{U}}^{\rm TL}(t) = C_{\rm ref}(t + \delta t) [C_{\rm ref}(t)]^{-1}.
\end{equation}
We then identify the lifetime $\tau_{\mathcal{U}} $ of the TL generator $\bm{\mathcal{U}}(t) $ and subsequently identify the characteristic memory distance $ d_{\mathcal{U}}$ using the following protocol:
\begin{enumerate}

    \item We generate TL-GME dynamics (Eq.~\ref{eq:tcl-GME}) using various candidate lifetimes, $\tau$. Choosing a lifetime implies that we treat all elements of $ \bm{\mathcal{U}}(t) $ as constant for $t \geq \tau$. We then compute the error between the exact dynamics and those generated by the TL-GME and quantify the error using the $ ||\text{L}||_2$ norm, i.e., $\mathrm{Error} = \frac{1}{N_t} ||C_{\rm HEOM}(t) - C_{\rm GME}(t)||_2$. Here $N_t$ is the number of time points for which GME predicts the dynamics. By plotting the error against the chosen lifetimes, we look for a plateau in the graph as it approaches zero. The time this occurs indicates the lifetime $ \tau_{\mathcal{U}}$ of the generator. Our threshold for an acceptable error value per element of $\bm{C}(t)$ is $1 \times 10^{-7}$. See Fig.~S1 in the supplementary information.
 
    \item Similarly, we employ the generator $ \bm{\mathcal{U}}^{\rm TL}(t)$ and construct the space-local (SL) GME dynamics with different characteristic memory distance ${d}_{\mc{U}}$. Choosing a specific memory distance means we set all the elements in the generator beyond that range set to $0$. Similarly, we compute the error between the exact dynamics and those generated by SL-GME with different $d_{\mc{U}}$ and plot the error as a function of the proposed cutoff distance. The distance at which the error minimizes we choose as our characteristic memory distance $d_{\mc{U}}$. See Fig.~S2 in the supplementary information.
    
\end{enumerate}

For an $N$-site lattice in 1D, finding that $d_{\mathcal{U}} < N/2$ ensures that we can use the dynamics arising from this $N$-site lattice as our reference. We can extend the size of the generator by leveraging translational invariance, transforming it from an $N \times N \times N_{\rm tsteps}$ tensor to an $M \times M \times N_{\rm tsteps}$ tensor, where $M > N$, by populating the additional entries with zeros. Here, $N_{\rm tsteps}$ is the number of timesteps in the original $\bm{\mathcal{U}}^{\rm TL}$ tensor. This augmentation procedure leads to the STL generator, $\bm{\mc{U}}^{\rm STL}$.

We utilize the full generator, $\bm{\mc{U}}^{\rm STL}(t)$, to generate the dynamics of the augmented system, $\bm{C} (t)$ with dimensions $M \times M$ at a given time $t$. Before reaching the generator lifetime, $\tau_{\mathcal{U}}$, one must use Eq.~\ref{eq:tcl-GME} for the time propagation. After $\tau_{\mathcal{U}}$, the generator becomes a constant matrix $\bm{\mathcal{U}} (\tau_{\mathcal{U}}) $. From this point, one propagate the dynamical $M\times M$ matrix $\bm{C} (t)$ to arbitrary times by simple matrix multiplication:
\begin{equation}\label{eq:tcl-gme-after-tau}
    \bm{C} (t + n \delta t) = [\bm{\mathcal{U}}(\tau_{\mathcal{U}})]^n \bm{C}(t = \tau_{\mathcal{U}}).
\end{equation}

\vspace{-8pt}
\subsection{Conserving population in GME} \label{population-cons}
\vspace{-5pt}

In our STL-GME approach, we truncate elements of $ \mathcal{U} $ that connect sites separated by distances greater than the cutoff distance, $ d_{\mathcal{U}}$. However, when one truncates such distant elements that contribute only a small amount to the total generator, it is possible to subtly break the generator's sum rules that ensure probability conservation. This is akin to the common practice in tensor-based algorithms like time-dependent density matrix renormalization group and time-evolving block decimation~\cite{schollwock_density-matrix_2011}, where one neglects singular values after SVD decomposition of the dynamical quantities, e.g., wavefunctions, below an acceptable threshold. Like in tensor-based methods, we employ one of two renormalization protocols to conserve probability:

\begin{enumerate}
    \item Method 1: Redistribute the discarded elements among the remaining elements after truncation. This can be mathematically expressed as:
    \begin{align}
    \mathcal{U}[i,j; d < d_{\mathcal{U}}] = & \: \mathcal{U}[i,j; d < d_{\mathcal{U}}] \nonumber \\
    & + \frac{1}{N_{d_{\mathcal{U}}}} \sum_j \mathcal{U}[i,j; d > d_{\mathcal{U}}], \quad \forall i,
    \end{align}
    where $ N_{d_{\mathcal{U}}} $ is the number of elements in $ \bm{\mathcal{U}} $ that remain after spatial truncation.

    \item Method 2: Renormalize the generator $\bm{\mathcal{U}} $ at each time step following spatial truncation:
    \begin{equation} \label{eq:renormalization}
    \mathcal{U}[i,j; d < d_{\mathcal{U}}] = \frac{\mathcal{U}[i,j; d < d_{\mathcal{U}}]}{\sum_j \mathcal{U}[i,j; d < d_{\mathcal{U}}]}, \quad \forall i.
    \end{equation}
\end{enumerate}

For 1D simulations, we employed Method 1. We used Method 2 in 2D simulations due to its superior ability to conserve population. 

\vspace{-8pt}
\subsection{GME in a 2D lattice} \label{2d-gme}
\vspace{-5pt}

For homogeneous lattices, we can perform just one simulation and leverage translational invariance to populate the entire correlation matrix $ \bm{C} (t) $. This approach applies to both 1D and 2D cases. However, for a periodic lattice with two distinct sites, we need to perform two separate HEOM simulations: one where the charge is initiated at the blue site and another where it is initiated at the red site (as shown in Fig.~\ref{fig:periodic_lattice}). One can then leverage translational invariance to populate the $ \bm{C} (t)$. Similarly, for a periodic lattice with $ M $ distinct lattice sites, one must perform $ M $ different simulations to construct the reference $\bm{C} (t) $.

To construct the generator for the size-expanded STL-GME, we use $C_{\rm ref}(t)$ calculated using the numerically exact HEOM solver. Before outlining the STL-GME procedure, we note that the time-local generator is a 3-indexed tensor, $\mc{U}[N,N, tsteps]$, where $N = N_x \times N_y$ denotes the total size of the lattice. The index that runs over the total sites of the lattice, $\alpha$, collapses the indices that label the $x$ and $y$ directions, $i$ and $j$, respectively 
\begin{equation}\label{eq:index-collapse-2D}
    \alpha = i + N_x \times j.  
\end{equation}
Having set this notation, we the steps for the STL-GME construction are as follows:
\begin{enumerate}

    \item Construct $ \bm{\mathcal{U}}^{\rm TL}(t)$ from $C_{\rm ref}(t)$ using Eq.~\ref{def-u} in the dimension of the reference calculation.

    \item Identify the lifetime $\tau_{\mc{U}}$ by following the procedure in Methods Sec.~\ref{GME-framework}.

    \item Identify the characteristic distance, $d_{\mathcal{U}}$ using the procedure outlined in step 2 in Methods Sec.~\ref{GME-framework}. To do this, see below for the protocol on choosing a distance cutoff $d_{\mathcal{U}}$ by truncating elements of the generator that connect sites beyond $d_{\mathcal{U}}$. 

    \item Reshape the generator matrix $\mc{U}[N, N; t]$ at all times up to $\tau_{\mc{U}}$ from a representation where $x$ and $y$ indices are collapsed onto a single index to a 5-tensor, where each dimension spans its original indices, $\mc{U}_{\tau_{\mc{U}}}[N, N; t] \rightarrow \mc{U}_{\tau_{\mc{U}}}[N_x, N_y; N_x, N_y; t]$. 

    \item For a homogeneous  2D lattice, consider the $N_x \times N_y$ submatrix defined by focusing on a particular initial excitation position (i.e., the second indices in the 5-tensor), say the 2D grid origin, $\mc{U}_{\tau_{\mc{U}}} [N_x, N_y; 0, 0; t]$ to implement spatial truncation. 

    \item Implement the spatial truncation based on the characteristic distance, $d_{\mc{U}}$. For the $\mc{U}_{\tau_{\mc{U}}} [N_x, N_y; 0, 0; t]$ submatrix, calculate the distance $d$ between the initial point $(0,0)$ and all sites ($i,j$) where carriers are measured as 
    \begin{equation}
        d = \sqrt{(i^2 + j^2 )} \quad \forall\ i,j.
    \end{equation}
    Then, set all elements of the generator submatrix where $d> d_{\mc{U}}$ to as $0$:
    \begin{equation}
       \mathcal{U}[i,j,0,0;t; d < d_{\mathcal{U}}] = 0 \quad \forall\ i,j.
    \end{equation}
 
    \item Apply the norm-conserving renormalization scheme in Eq.~\ref{eq:renormalization} to the generator elements $\mc{U}_{\tau_{\mc{U}}} [N_x,N_y;0,0;t]$ to build $\mc{\Tilde{U}}_{\tau_{\mc{U}}}[N_x,N_y;0,0;t]$. 
    
    \item Exploit translational symmetry to populate the modified generator $\mc{\Tilde{U}}_{\tau_{\mc{U}}} [N_x,N_y; N_x,N_y;t]$ by iterating steps 4-5 over all initial conditions, $(k,l)$ in $\mc{\Tilde{U}}_ {\tau_{\mc{U}}} [N_x,N_y; k,l;t]$. 

    \item Reshape the 5-tensor generator after spatial truncation into matrix $\mc{\Tilde{U}}_{\tau_{\mc{U}}} [N, N, t]$ by collapsing the $x$ and $y$ indices of the initial and final conditions to a compound single index as dictated by Eq.~\ref{eq:index-collapse-2D}. 
    
    \item Employ $\mc{\Tilde{U}} [N, N; t]$ to generate 2D-GME dynamics. As in the 1D case, one can use Eq.~\ref{eq:tcl-GME} for time $t\leq\tau_{\mc{U}}$ and Eq.~\ref{eq:tcl-gme-after-tau} for $t>\tau_{\mc{U}}$ to propagate $C_{\mathrm{GME}} (t)$.

    \item Calculate the error between $C_{\mathrm{GME}} (t)$ and $C_{\mathrm{ref}} (t)$ using the procedure in Methods Sec.~\ref{GME-framework} for a range of proposed distance cutoffs.
    
    \item Plot the $||\mathrm{L}||_2$ error as a function of the distance cutoff to identify the characteristic memory distance, $d_{\mc{U}}$, where the error converges to a predefined threshold (similar to Fig.~S2 in the supplementary information).
    
\end{enumerate}

Following these previous steps, one can construct  $C_{\mathrm{GME}} (t)$ for the \textit{same system size} as the reference calculation. To obtain the dynamics of \textit{extended system size}, one needs to follow the following steps:

\begin{enumerate}
    \item Augment the dimension of the generator. Start with the norm-conserving generator in reshaped space, $\mc{\Tilde{U}}[N_x,N_y;0,0; t]$ for all times $t \in [0,\tau_{\mathcal{U}}]$ (see step 7 of the previous protocol). Expand the size of each spatial dimension, taking $N_x \rightarrow M_x$ and $N_y \rightarrow M_y$ and add zeros to the new entries of the tensor. This makes new extended element $\mc{\Tilde{U}^{\rm ex}}[M_x,M_y;M_x,M_y; t]$, where $M_x \ge N_x$ and $M_y \ge N_y$.

    \item Following steps 8 and 9 of the previous protocol to construct the extended generator $\mc{\Tilde{U}^{\rm ex}}[M_x,M_y;M_x,M_y; t] \to \mc{\Tilde{U}^{\rm ex}}_{\tau_{\mc{U}}} [M, M; t]$, where $M = M_x \times M_y$.

    \item Employ the generator $\mc{\Tilde{U}^{\rm ex}}[M, M; t]$ to construct $C_{\rm GME} (t)$ for extended $M \times M$ lattice (see step 10 in the above protocol).
\end{enumerate}

For the 2D homogeneous lattice in Fig.~\ref{fig:2dvs1d_polaron}, we employ the polaron dynamics on an $8 \times 8$-site lattice as the reference simulation to predict the dynamics of a $30 \times 30$ lattice system. 
 
For the periodic 2D lattice with two distinct lattice points (see Fig.~\ref{fig:periodic_lattice}), we perform two reference simulations on a $10 \times 10$-site lattice, one starting from red site and another form blue site. To construct $C_{\rm ref} (t)$, which is a $[N, N, tsteps]$-dimensional tensor, we break the lattice index into the $x$ and $y$ coordinate indices using Eq.~\ref{eq:index-collapse-2D}. We populate $C_{\rm ref} [N_x, N_y; 0, 0; t]$ with reference simulation initialized at the blue site and $C_{\rm ref} [N_x, N_y; 0, 1; t]$ with reference results from initialization at red site. We then collapse the second coordinate indices of $C_{\rm ref} [N_x, N_y; k, l;t]$  using Eq.~\ref{eq:index-collapse-2D} such that $\beta = k + N_x \times l.$ For all $k,l$, if $\beta$ is even we use $C_{\rm ref} [N_x, N_y; 0, 0; t]$ and if $\beta$ is odd we employ $C_{\rm ref} [N_x, N_y; 0, 1; t]$ by invoking translational invariance to populate $C_{\rm ref} [N_x, N_y; N_x, N_y; t]$. After that, we collapse the coordinate indices into lattice indices to construct $C_{\rm ref} [N, N]$ for our GME. As in the homogeneous case, we follow the rest of the protocol unchanged, except at steps 6-8. At these steps, one needs to calculate the distance now for $\mc{U}_{\tau_{\mc{U}}} [N_x,N_y;0,0;t]$ and $\mc{U}_{\tau_{\mc{U}}} [N_x,N_y;0,1;t]$ separately. In principle, one can implement separate distance cutoff in each case. Once one has implemented the distance cutoff and norm-conservation schemes, we again employ odd-even rules like those in $C_{\rm ref} (t)$ to construct the generator $\mc{\Tilde{U}} [N_x,N_y; N_x,N_y; t].$ All other steps remain unchanged. To extend the system size, one needs to follow same odd-even protocol in the augmentation of the generator.

\onecolumngrid

\vfill\pagebreak

\setcounter{section}{0}
\setcounter{equation}{0}
\setcounter{figure}{0}
\setcounter{table}{0}
\setcounter{page}{1}

\renewcommand{\thepage}{S\arabic{page}}
\renewcommand{\theequation}{S\arabic{equation}}
\renewcommand{\thefigure}{S\arabic{figure}}

\section*{Supplemental information: Nonequilibrium relaxation exponentially delays the onset of quantum diffusion}

\section{Lifetime identification}

\begin{figure}[ht]
\begin{center} 
\vspace{-3pt}
    \resizebox{.5\textwidth}{!}{\includegraphics[trim={00pt 0pt 0pt 0pt},clip]{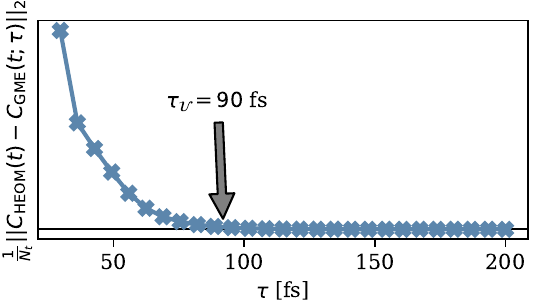}}
    \vspace{-15pt}
\end{center}
\caption{\label{fig:lifetime} The error between GME and exact HEOM dynamics as a function of lifetime candidates $\tau$ reveals the lifetime $\tau_{\mathcal{U}} = 90$~fs. This result is for the 1D dispersive Holstein chain with parameters $v = 322$~cm$^{-1}$, $\eta = 2984$~cm$^{-1}$, $\gamma = 400$~cm$^{-1}$, and $T = 300 $~K. The black horizontal line indicates our threshold of $1 \times 10^{-7}.$}
\vspace{-5pt}
\end{figure}

In the Method section, we discuss how to identify the generator lifetime, $\tau_{\mathcal{U}}$. Here, Fig.~\ref{fig:lifetime} shows the convergence of the error metric $\frac{1}{N_t} ||C_{\rm HEOM}(t) - C_{\rm GME}(t)||_2$ between GME and HEOM dynamics as a function of proposed lifetime, $\tau$. Our convergence threshold per element of the correlation matrix is $1\times 10^{-7}$. Thus, we find $\tau_{\mathcal{U}} = 90$~fs.

\section{characteristic memory distance identification}

\begin{figure}[ht]
\begin{center} 
\vspace{-3pt}
    \resizebox{.5\textwidth}{!}{\includegraphics[trim={00pt 0pt 0pt 0pt},clip]{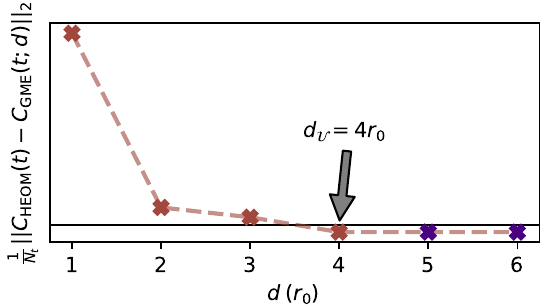}}
    \vspace{-15pt}
\end{center}
\caption{\label{fig:du} The error between GME and exact HEOM dynamics as a function of distance cutoffs, $d$, reveals that $d_{\mathcal{U}} = 4 r_0$. This result is for the 1D dispersive Holstein chain with parameters $v = 322$~cm$^{-1}$, $\eta = 2984$~cm$^{-1}$, $\gamma = 700$~cm$^{-1}$, and $T = 300 $~K. The black horizontal line indicates our threshold. }
\vspace{-5pt}
\end{figure}

In the Method section, we discuss how to identify characteristic memory distance $d_{\mathcal{U}}$. To find $d_{\mathcal{U}}$, we compute the GME dynamics for a set of proposed memory distances, $d$. Here, Fig.~\ref{fig:du} shows the convergence of the error metric $\frac{1}{N_t} ||C_{\rm HEOM}(t) - C_{\rm GME}(t)||_2$ between GME and HEOM dynamics as a function of proposed distance cutoff, $d$. We first employ a 10-site to compute the error (brown crosses) to find $d_{\mathcal{U}} = 4 r_0$. To confirm the cutoff, we compare to results from a 30-site simulation, where the error lies below our threshold, even when we increase the cutoff distance more than $4 r_0$ (purple crosses). Our error threshold per element is $5 \times 10^{-6}$.

\section{Validity of the 1D outer product construction}

\begin{figure}[!ht]
\begin{center} 
\vspace{-3pt}
    \resizebox{.9\textwidth}{!}{\includegraphics[trim={00pt 0pt 0pt 0pt},clip]{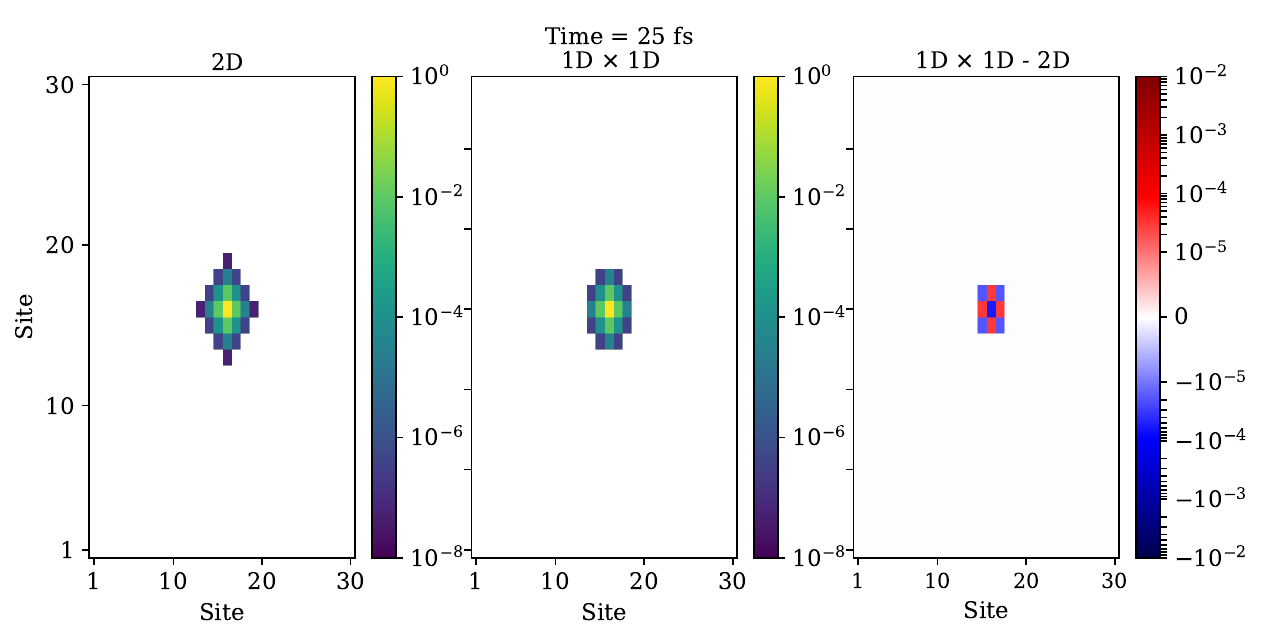}}
    \vspace{-15pt}
\end{center}
\caption{\label{fig:t25fs} Polaron spread in a $30 \times 30$-site 2D lattice, a 1D $30$-site lattice outer product, and the difference between the two at $t = 25$~fs. Results are for the parameter regime of the dispersive Holstein lattice used in Table.~\ref{table1}.}
\vspace{-5pt}
\end{figure}

\begin{figure}[!ht]
\begin{center} 
\vspace{-3pt}
    \resizebox{.9\textwidth}{!}{\includegraphics[trim={00pt 0pt 0pt 0pt},clip]{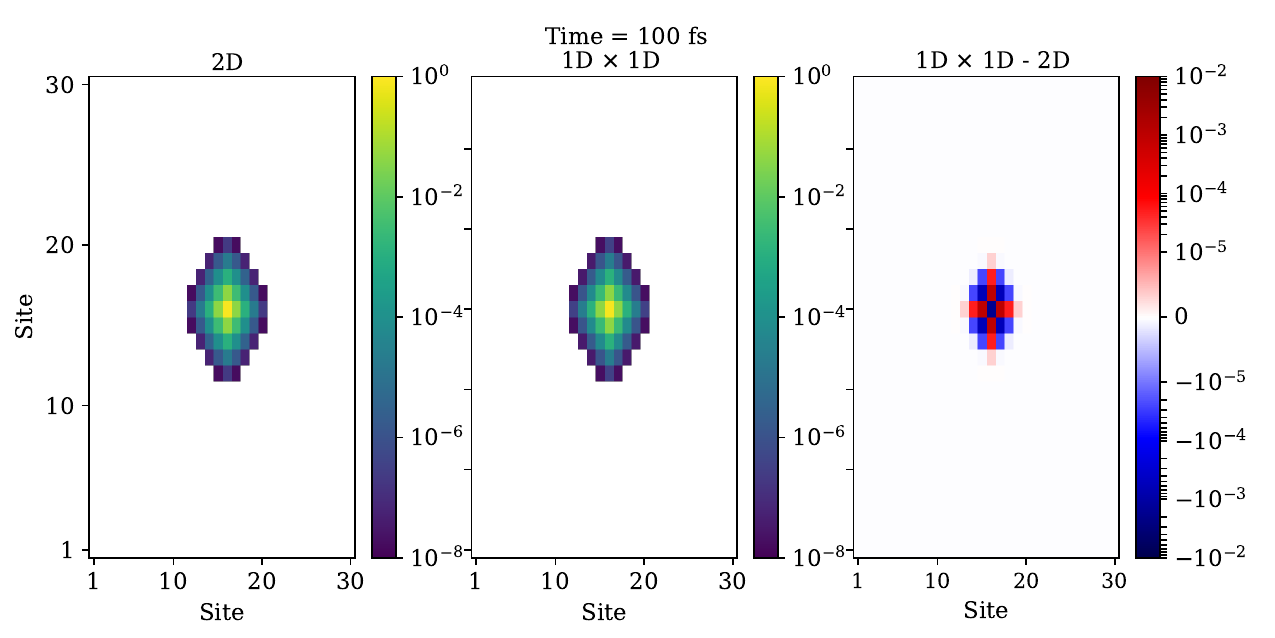}}
\end{center}
\caption{\label{fig:t100fs} Polaron spread in a $30 \times 30$-site 2D lattice, a 1D $30$-site lattice outer product, and the difference between the two at $t = 100$~fs. Results are for the parameter regime of the dispersive Holstein lattice used in Table.~\ref{table1}.}
\vspace{-5pt}
\end{figure}

\begin{figure}[!ht]
\begin{center} 
\vspace{-3pt}
    \resizebox{.9\textwidth}{!}{\includegraphics[trim={00pt 0pt 0pt 0pt},clip]{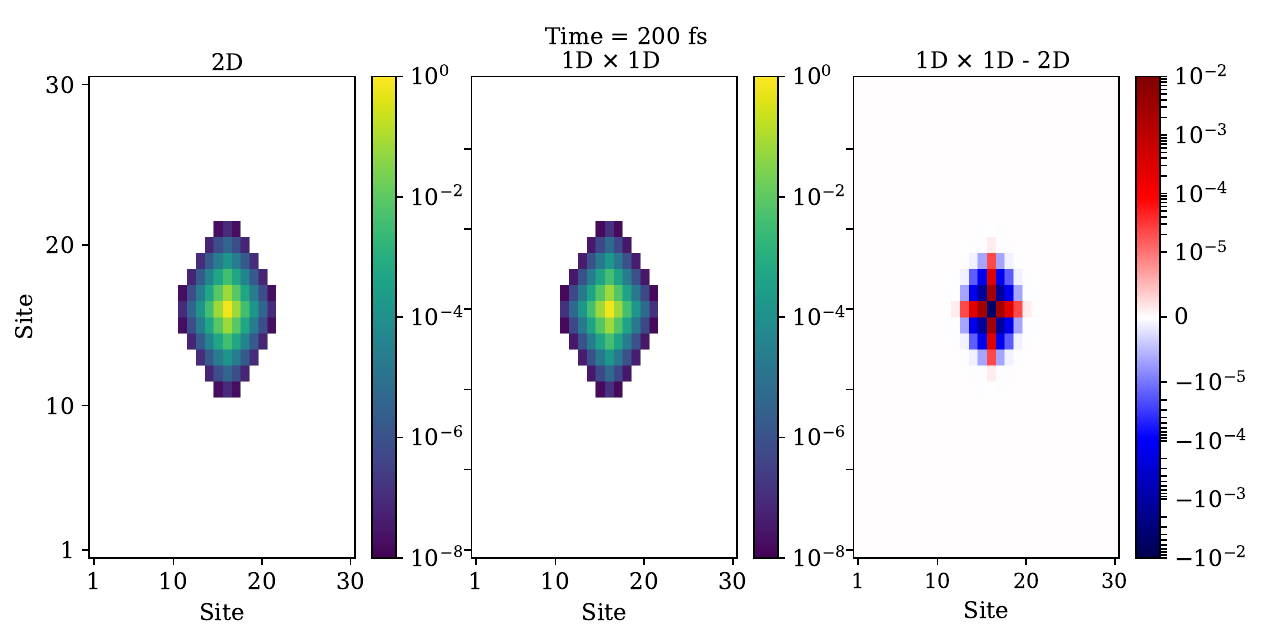}}
    \vspace{-15pt}
\end{center}
\caption{\label{fig:t200fs} PPolaron spread in a $30 \times 30$-site 2D lattice, a 1D $30$-site lattice outer product, and the difference between the two at $t = 200$~fs. Results are for the parameter regime of the dispersive Holstein lattice used in Table.~\ref{table1}.}
\vspace{-5pt}
\end{figure}

\begin{figure}[!ht]
\begin{center} 
\vspace{-3pt}
    \resizebox{.9\textwidth}{!}{\includegraphics[trim={00pt 0pt 0pt 0pt},clip]{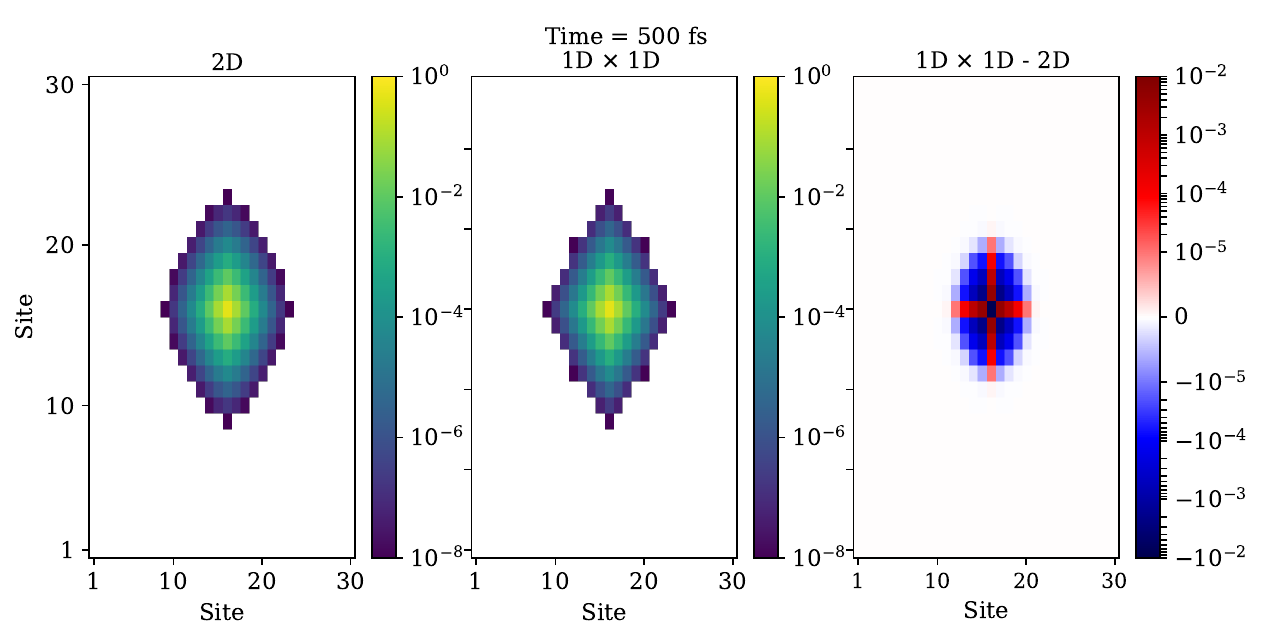}}
    \vspace{-15pt}
\end{center}
\caption{\label{fig:t500fs} Polaron spread in a $30 \times 30$-site 2D lattice, a 1D $30$-site lattice outer product, and the difference between the two at $t = 500$~fs. Results are for the parameter regime of the dispersive Holstein lattice used in Table.~\ref{table1}.}
\vspace{-5pt}
\end{figure}

\begin{figure}[!ht]
\begin{center} 
\vspace{-3pt}
    \resizebox{.9\textwidth}{!}{\includegraphics[trim={00pt 0pt 0pt 0pt},clip]{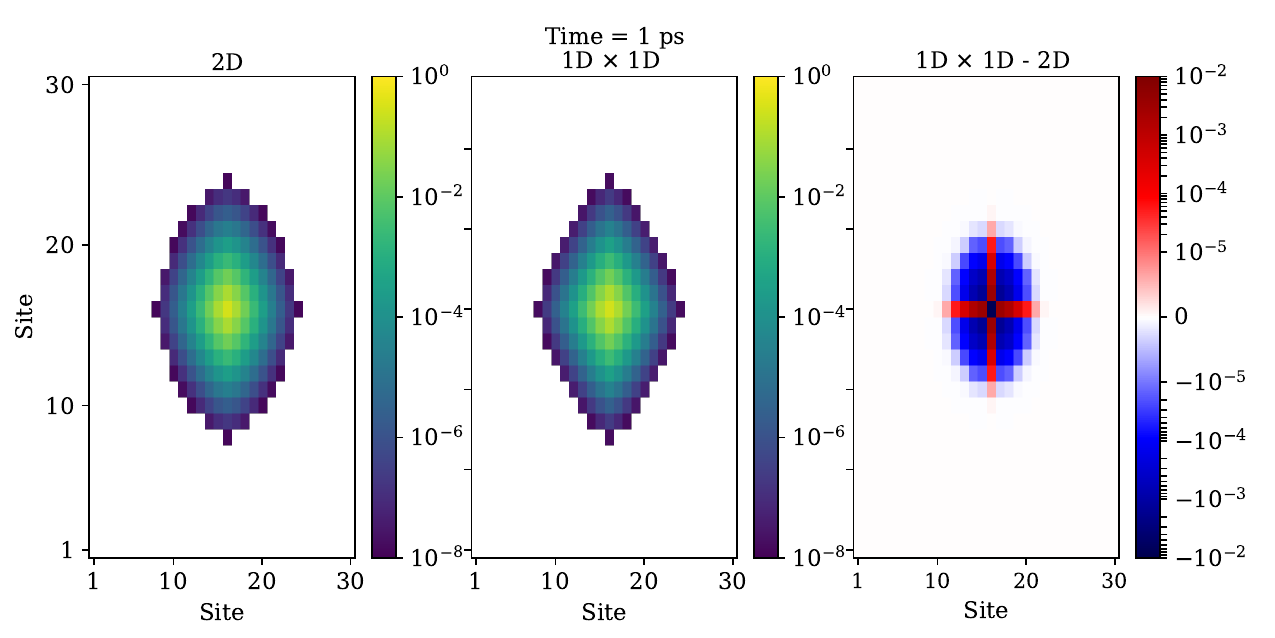}}
    \vspace{-15pt}
\end{center}
\caption{\label{fig:t1ps} Polaron spread in a $30 \times 30$-site 2D lattice, a 1D $30$-site lattice outer product, and the difference between the two at $t = 1$~ps. Results are for the parameter regime of the dispersive Holstein lattice used in Table.~\ref{table1}.}
\vspace{-5pt}
\end{figure}

\begin{figure}[!ht]
\begin{center} 
\vspace{-3pt}
    \resizebox{.9\textwidth}{!}{\includegraphics[trim={00pt 0pt 0pt 0pt},clip]{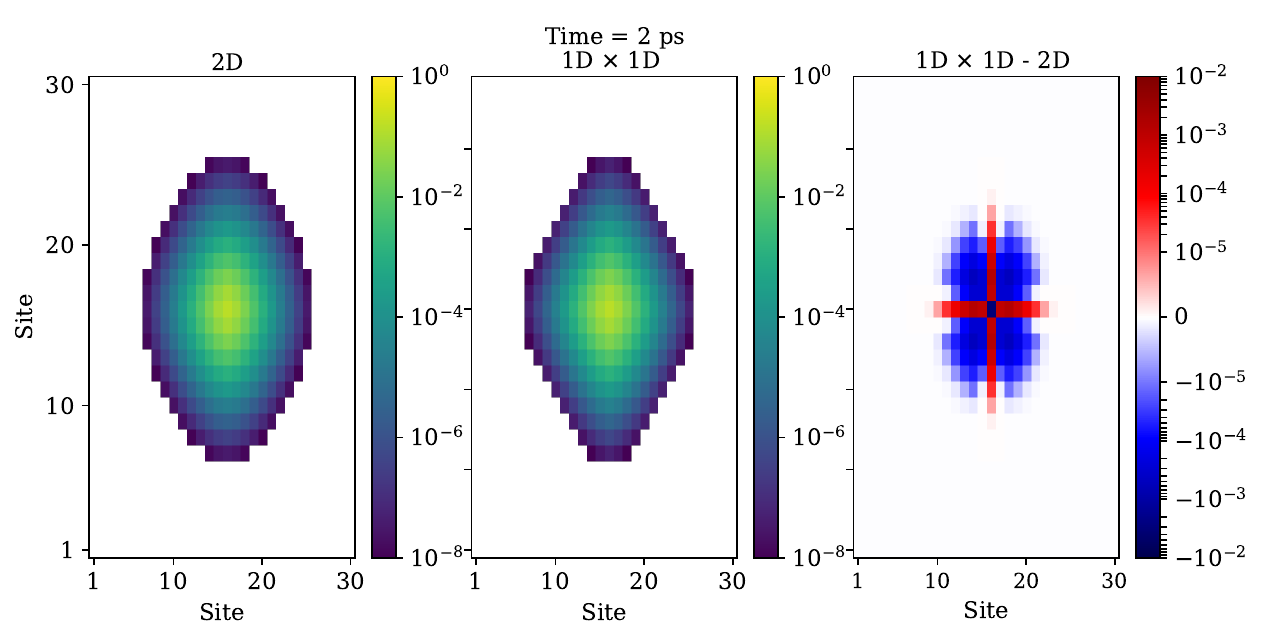}}
    \vspace{-15pt}
\end{center}
\caption{\label{fig:t2ps} Polaron spread in a $30 \times 30$-site 2D lattice, a 1D $30$-site lattice outer product, and the difference between the two at $t = 2$~ps. Results are for the parameter regime of the dispersive Holstein lattice used in Table.~\ref{table1}.}
\vspace{-5pt}
\end{figure}

\begin{figure}[!ht]
\begin{center} 
\vspace{-3pt}
    \resizebox{.9\textwidth}{!}{\includegraphics[trim={00pt 0pt 0pt 0pt},clip]{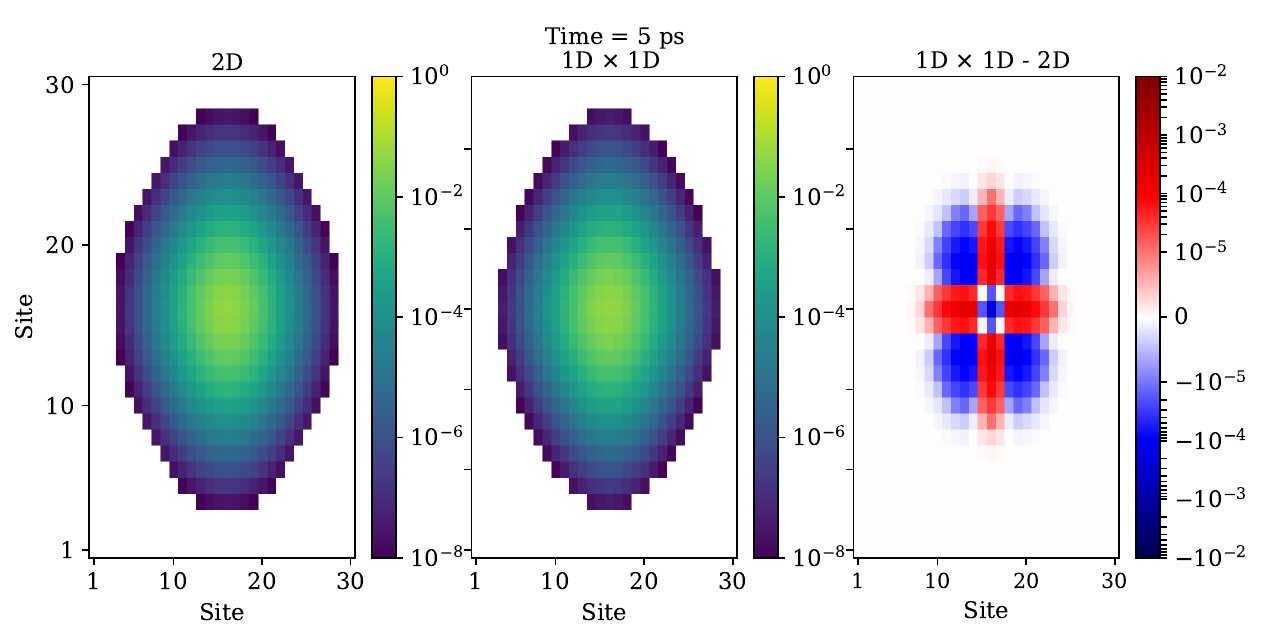}}
    \vspace{-15pt}
\end{center}
\caption{\label{fig:t5ps} Polaron spread in a $30 \times 30$-site 2D lattice, a 1D $30$-site lattice outer product, and the difference between the two at $t = 5$~ps. Results are for the parameter regime of the dispersive Holstein lattice used in Table.~\ref{table1}.}
\vspace{-5pt}
\end{figure}

\begin{figure}[!ht]
\begin{center} 
\vspace{-3pt}
    \resizebox{.9\textwidth}{!}{\includegraphics[trim={00pt 0pt 0pt 0pt},clip]{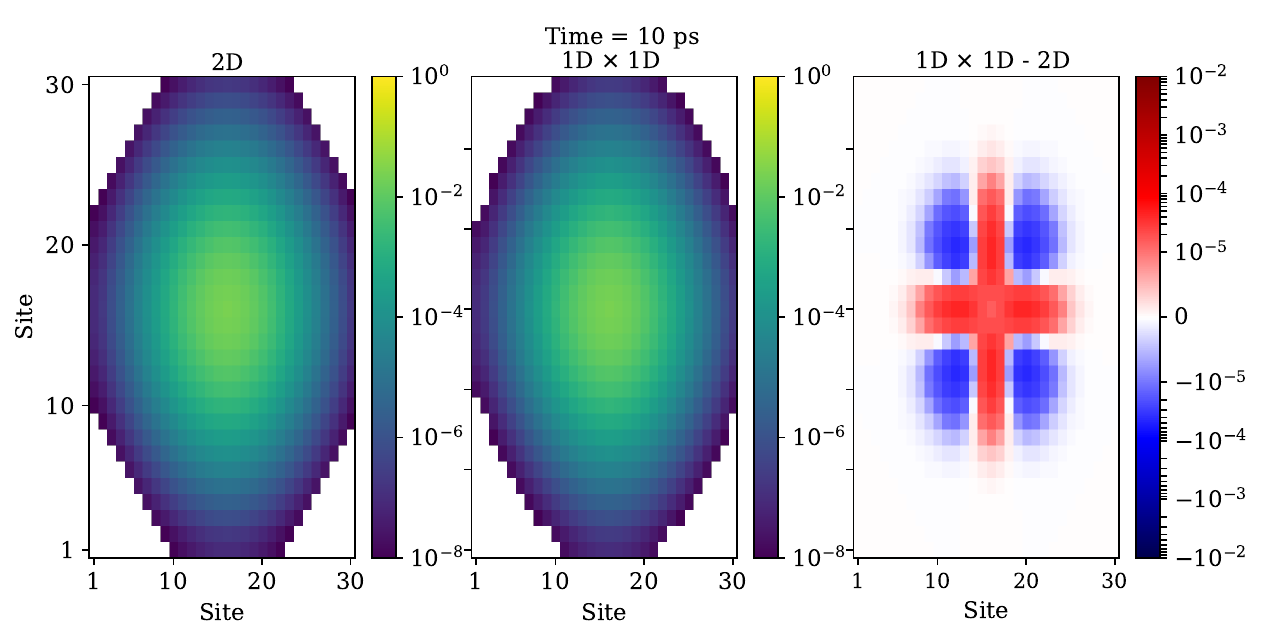}}
    \vspace{-15pt}
\end{center}
\caption{\label{fig:t10ps} Polaron spread in a $30 \times 30$-site 2D lattice, a 1D $30$-site lattice outer product, and the difference between the two at $t = 10$~ps. Results are for the parameter regime of the dispersive Holstein lattice used in Table.~\ref{table1}.}
\vspace{-5pt}
\end{figure}

\begin{figure}[!ht]
\begin{center} 
\vspace{-3pt}
    \resizebox{.9\textwidth}{!}{\includegraphics[trim={00pt 0pt 0pt 0pt},clip]{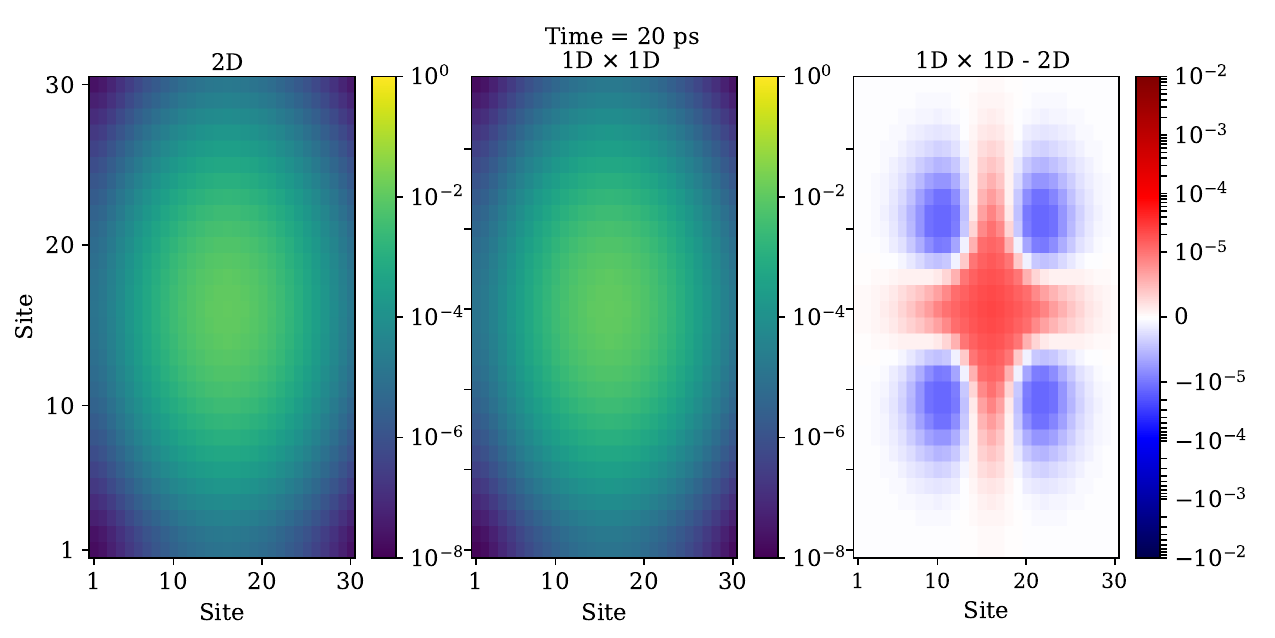}}
    \vspace{-15pt}
\end{center}
\caption{\label{fig:t20ps} Polaron spread in a $30 \times 30$-site 2D lattice, a 1D $30$-site lattice outer product, and the difference between the two at $t = 20$~ps. Results are for the parameter regime of the dispersive Holstein lattice used in Table.~\ref{table1}.}
\vspace{-5pt}
\end{figure}

\begin{figure}[!ht]
\begin{center} 
\vspace{-3pt}
    \resizebox{.9\textwidth}{!}{\includegraphics[trim={00pt 0pt 0pt 0pt},clip]{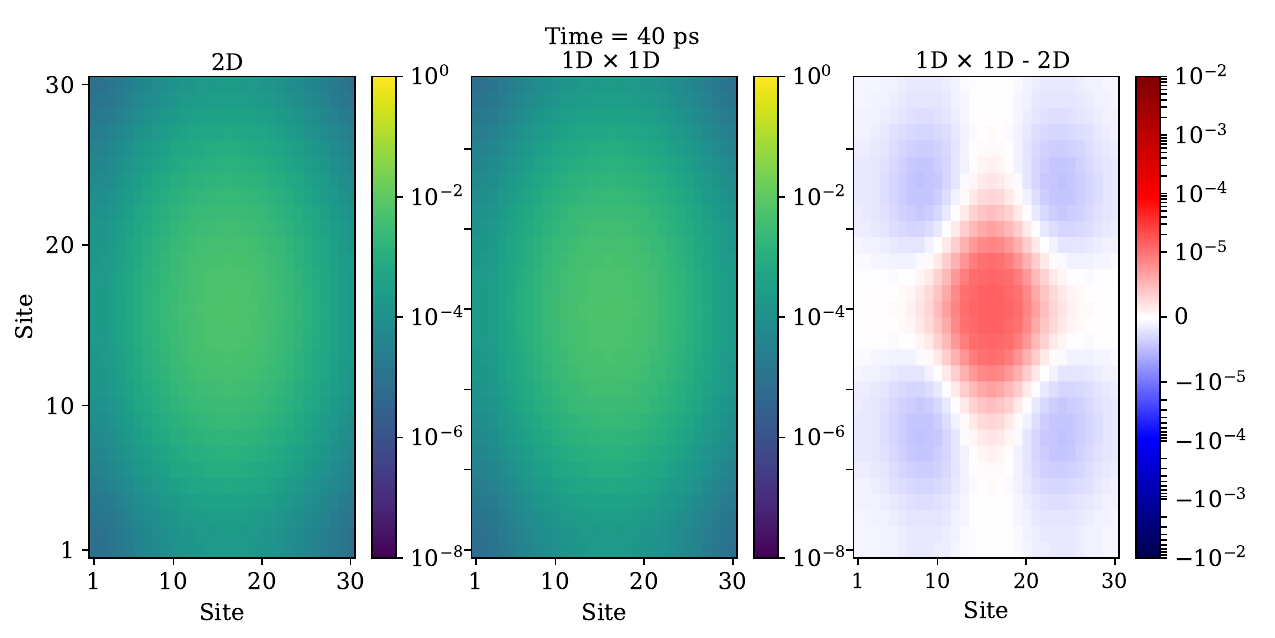}}
    \vspace{-15pt}
\end{center}
\caption{\label{fig:t40ps} Polaron spread in a $30 \times 30$-site 2D lattice, a 1D $30$-site lattice outer product, and the difference between the two at $t = 40$~ps. Results are for the parameter regime of the dispersive Holstein lattice used in Table.~\ref{table1}.}
\vspace{-5pt}
\end{figure}

In the main text, we ask if constructing the outer product of polaron spread in 1D can capture the polaron spared in 2D. Figures~\ref{fig:t25fs} -~\ref{fig:t40ps} show that while 1D outer products and 2D polaron densities differ initially, the deviation diminishes over time. We show the plots for $10$ time instances. In Table.~\ref{table1}, we record the maximum of the absolute value of the difference between the polaron population of any site between the direct 2D simulation and the 1D outer product.

\begin{table}[!ht]
    \centering
    \vspace{0pt}
    \hspace{0pt}
    \begin{tabular}{|c|c|} 
    \hline
    \qquad  Time \qquad \qquad  &  Max.~of absolute value of the difference \\
     & in polaron density between 2D $\&$ 1D outer product \\
     \hline
     25 fs & $5.4 \times 10^{-5}$  \\ \hline
     100 fs & $1.6 \times 10^{-3}$   \\ \hline
     200 fs & $4.2 \times 10^{-3}$  \\ \hline
     500 fs & $8.2 \times 10^{-3}$   \\ \hline
     1 ps & $7.6 \times 10^{-3}$    \\ \hline
     2 ps & $3.2 \times 10^{-3}$  \\ \hline
     5 ps & $2.1 \times 10^{-4}$   \\ \hline
     10 ps & $4.5 \times 10^{-5}$   \\ \hline
     20 ps & $2.4 \times 10^{-5}$   \\ \hline
    40 ps &  $1.5 \times 10^{-5}$    \\ \hline
    \end{tabular} 
    \vspace{-4pt}
    \caption{\label{table1} Maximum of the absolute value of the difference between population correlation matrix of the 2D simulation versus the 1D outer product for the dispersive Holstein parameters $v = 25$~cm$^{-1}$, $\eta = 500$~cm$^{-1}$, $\gamma = 41$~cm$^{-1}$, and $T = 300 $~K. }
    \vspace{-0pt}
\end{table}

\section{Polaron motion in 2D periodic lattices}

In Fig.~4 of the main text, we illustrate the spread of the polaron in a periodic lattice with two distinct lattice points at $t \approx 40$~ps considering two different energy scale variations. The accompanying \href{https://o365coloradoedu-my.sharepoint.com/:v:/g/personal/srbh4687_colorado_edu/EXSjW8oWVQJPhJ6CqTA6wIABkmh03MpPe7JjPXxM84zSHQ?e=Z22fmn}{supporting video} provides a dynamic view of the polaron’s evolution in this lattice over time, extending up to $\sim120$~ps. The video demonstrates that when $\epsilon_1 > \epsilon_2$ and $v_1 > v_2 > v_3$, the polaron density remains primarily localized along the vertical axis. However, when $v_1 > v_3 > v_2$, the polaron density spreads elliptically and approaches equilibrium by $\sim120$~ps.
\end{document}